\documentclass[aps,prl,twocolumn,english,superscriptaddress,longbibliography]{revtex4-2}
\usepackage{amsmath,amssymb,amsfonts}
\usepackage{bm}
\usepackage{tikz}
\usepackage{babel}
\usepackage{color}
\usepackage{graphicx}
\usepackage{xspace}
\usepackage{epstopdf}
\usepackage{dcolumn}
\usepackage{longtable}
\usepackage{multirow}
\usepackage{physics}
\usepackage[colorlinks=true, letterpaper=true, pdfstartview=FitV, linkcolor=blue, citecolor=blue, urlcolor=blue]{hyperref}
\usepackage{dsfont}

\usepackage{mathrsfs}

\begin{document}

\title{Projective symmetry determined topology in flux Su-Schrieffer-Heeger model}

\author{Gang Jiang}
\affiliation{National Laboratory of Solid State Microstructures and Department of Physics, Nanjing University, Nanjing 210093, China}

\author{Z. Y. Chen}
\affiliation{National Laboratory of Solid State Microstructures and Department of Physics, Nanjing University, Nanjing 210093, China}

\author{S. J. Yue}
\affiliation{National Laboratory of Solid State Microstructures and Department of Physics, Nanjing University, Nanjing 210093, China}

\author{W. B. Rui}
\affiliation{Department of Physics and HKU-UCAS Joint Institute for Theoretical
	and Computational Physics at Hong Kong, The University of Hong Kong,
	Pokfulam Road, Hong Kong, China}
\affiliation{HK Institute of Quantum Science \& Technology, The University of Hong Kong, Pokfulam Road, Hong Kong, China}

\author{Xiao-Ming Zhu}
\affiliation{Hangzhou Yingshi Technology Co., Ltd, 153 Lianchuang Street, Yuhang District, Hangzhou 310000, China}

\author{Shengyuan A. Yang}
\affiliation{Research Laboratory for Quantum Materials, IAPME, University of Macau, Macau, China}

\author{Y. X. Zhao}
\email[]{yuxinphy@hku.hk}
\affiliation{Department of Physics and HKU-UCAS Joint Institute for Theoretical
	and Computational Physics at Hong Kong, The University of Hong Kong,
	Pokfulam Road, Hong Kong, China}
\affiliation{HK Institute of Quantum Science \& Technology, The University of Hong Kong, Pokfulam Road, Hong Kong, China}
\maketitle

\textbf{In the field of symmetry-protected topological phases, a common wisdom is that the symmetries fix the
topological classifications, but they alone cannot determine whether a system is topologically trivial or not. Here, we show that this is no longer true in cases where symmetries are projectively represented. Particularly, the Zak phase, a topological invariant of a one-dimensional system, can be entirely determined by the projective symmetry algebra (PSA). To demonstrate this remarkable effect, we propose a minimal model, termed as flux Su-Schrieffer-Heeger (SSH) model, where the bond dimerization in the original SSH model is replaced by a flux dimerization. We present experimental realization of our flux SSH model in an electric-circuit array, and our predictions are directly confirmed by experimental measurement. Our work refreshes the understanding of the relation between symmetry and topology, opens up new avenues for exploring PSA determined topological phases, and suggests flux dimerization as a novel approach for designing topological crystals.  }


The action of symmetries on a physical system is described by their representations. Such actions impose constraints on the physical state and decide its topological classification, i.e., what are the topologically distinct phases allowed by the symmetries.  Nevertheless,
knowing the symmetries and their representations does not automatically tell us which phase (trivial or nontrivial) the system is in, a fact well-known from previous studies~\cite{Kane_RMP,XLQi_RMP,Schnyder_RMP}.

\begin{figure}[t]
	\includegraphics[scale=1.0]{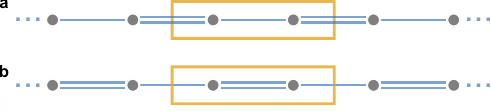}
	\caption{\textbf{Two phases of the standard SSH model. }  
	\textbf{a},~The topological phase with intercell bond stronger than the intra-cell bond. 
	\textbf{b},~The trivial phase with intercell bond weaker than the intra-cell bond. \label{fig:ssh}}
\end{figure}

Consider the famous SSH model~\cite{SSH}, as illustrated in Fig.~\ref{fig:ssh}. Topological classification of SSH model can be resulted from several choices of symmetries. Here, let's consider the spacetime inversion $PT$ being the protecting symmetry, which leads to a $\mathbb{Z}_2$ classification, with the nontrivial and trivial phases characterized by the Zak phase $\gamma=\pi$ and $0$, respectively~\cite{Zak_phase,Zhao_PT_cls}.
In the SSH chain, the two topological phases corresponds to the two bond dimerization patterns shown in Fig.~\ref{fig:ssh}. Namely, in the unit cell (compatible with boundary condition), the nontrivial (trivial) phase has intercell bond stronger (weaker) than the intracell bond. One can see that the symmetry determines the $\mathbb{Z}_2$ classification, but it cannot determine which phase an SSH chain belongs to. Indeed, the two phases in Fig.~\ref{fig:ssh} correspond to the same symmetry representation, with the same symmetry group relations
\begin{equation}\label{or}
  (PT)^2=1,\qquad (PT)L(PT)=L^{-1},
\end{equation}
where $L$ is the unit translation along the chain. Note that in equations (here and hereafter), the symmetry symbols means their corresponding operators (i.e., representation in a physical system).


Recently, the framework of ordinary crystal symmetry groups has been extended into PSAs~\cite{PhysRevB.102.161117, PT_Switch_PRL, Xue_prl_2022, Qiu_PRL_2022,PT_Expert_PRL,chen2023classification,Jonah_PRL_2023,ZhangChen_prl_2023}. That is, the successive action of two crystal symmetry operators $S_1$ and $S_2$ may be modified by an additional phase factor~\cite{Wigner_Projective}:
$\rho(S_1)\rho(S_2)=\Omega(S_1,S_2)\rho(S_1 S_2)$, where $\rho(S)$ denotes the representation of $S$.
In general, the phase factor $\Omega(S_1,S_2)$ is valued in $U(1)$, and in the presence of time-reversal symmetry $T$, it will be restricted to $\mathbb{Z}_2=\{\pm 1\}$~\cite{chen2023classification}.

In the framework of PSAs, a remarkable discovery is that the Zak phase $\gamma$ is an invariant of a PSA, i.e., it is completely determined by the symmetries' projective representations.

\begin{figure*}[t]
	\includegraphics[scale=1.0]{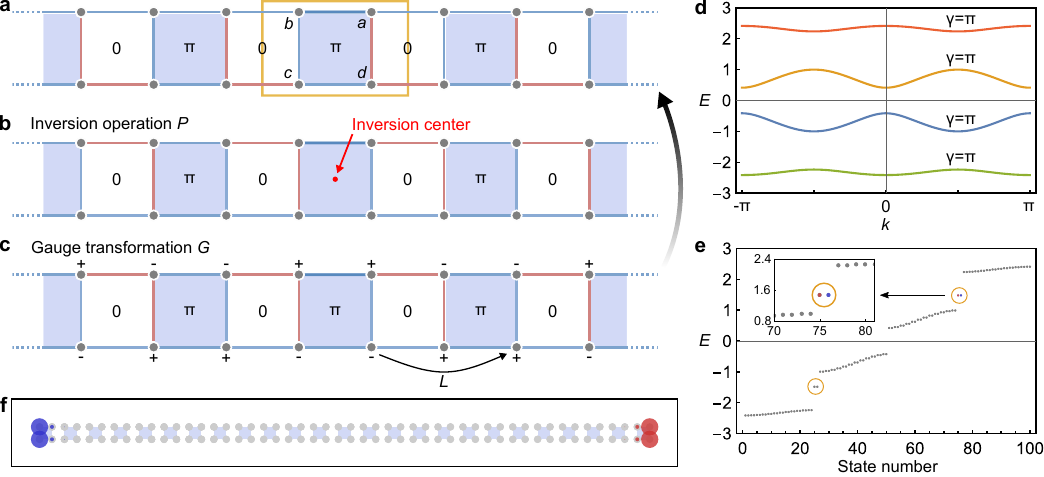}
	\caption{\textbf{The flux SSH model. }  
	\textbf{a},~Illustration of the flux SSH model. A unit cell (marked by the yellow box) contains four sites, labeled by a, b, c and d. All hopping amplitudes have the same magnitude but may have different signs. The negative and positive ones are marked in red and blue, respectively. These signs result in a flux distribution as indicated in the figure.  
	\textbf{b},~Direct inversion operation with respect to the inversion center (red dot) does not preserve the gauge configuration, i.e., the color of the bonds are changed from those in \textbf{a}. 
	\textbf{c},~An additional gauge transformation $G$ is required to recover the original configuration in \textbf{a}. Here, the plus and minus signs indicate the phase change of the local basis at respective sites. One observes that this $G$ does not commute with $L$ and $P$, as both  $L$ and $P$ exchange $\pm$ signs.
	\textbf{d},~Band structure of the flux SSH model. All bands have Zak phase $\pi$ as determined by the PSA.
	\textbf{e},~Energy spectrum for the model with a length of 25 unit cells. The in-gap states are surrounded by orange circles.  
	\textbf{f},~Spatial distribution of the two ingap states in the third gap as indicated in the insert of \textbf{e}. The two states are concentrated at two ends,  and therefore represent the topological end states corresponding to the nontrivial Zak phase. \label{fig:tb}}
\end{figure*}

To clarify this point, we introduce the flux SSH model, as shown in Fig.~\ref{fig:tb}\textbf{a}. In this model, a unit cell has four sites.
All hoppings between nearest neighbors have the \emph{same} magnitude, but their signs can be positive or negative, denoted by blue and red colors in Fig.~\ref{fig:tb}\textbf{a}. Going around a plaquette, the accumulation of hopping signs correspond to a gauge flux of $0$ or $\pi$. Our flux SSH model has an alternating distribution of $0$ and $\pi$ fluxes. In this sense, the bond dimerization in the original SSH model is replaced here by a `flux dimerization'.

Clearly, this model also preserves $P$, $T$, and $L$ symmetries. However, the flux dimerization modifies the representation of these symmetries and their PSA in a fundamental way. Consider the inversion operation. With the inversion center at the center of a $\pi$-flux plaquette, direct inversion transforms the chain from Fig.~\ref{fig:tb}\textbf{a} to Fig.~\ref{fig:tb}\textbf{b}. One observes that although the flux distribution is preserved,
the color of the bonds, i.e., the gauge connection configuration, is not. To recover the original configuration in Fig.~\ref{fig:tb}\textbf{a},
an additional gauge transformation $G$ is required. For our current case, $G$ is illustrated in Fig.~\ref{fig:tb}\textbf{c}, which involves
sign ($\pi$-phase) change at some of the local base states. It follows that the representation of inversion in the flux SSH model is a combined operation
\begin{equation}
  \mathbf{P}=GP.
\end{equation}
It is the combined operator $\mathbf{P}=GP$ that commutes with the Hamiltonian.

Importantly,  $G$ does not commute with $P$ and $L$. Instead, we have $PGP^{-1}=-G$ and $LGL^{-1}=-G$, as reflected in Fig.~\ref{fig:tb}\textbf{c} where both $P$ and $L$ inverse the sign configuration of $G$. Then, one immediately observes that the relations in (\ref{or}) satisfied by the ordinary representation are now modified into a PSA, with
\begin{equation}\label{PT2}
	(\mathbf{P}T)^2=-1,
\end{equation}
and
\begin{equation}\label{PTL}
  (\mathbf{P}T)L(\mathbf{\mathbf{P}}T)^{-1}=-L^{-1}.
\end{equation}
The minus signs in the two identities are invariants of the PSA as shown in Methods.

More importantly, the PSA in (\ref{PT2}) and (\ref{PTL}) dictates the value of Zak phase. To see this, note that in $k$ space, $L$ is represented by $e^{ik}$ (taking the lattice constant to be unit), then Eq.~(\ref{PTL}) indicates that
$\mathbf{P}T$ must send $k$ to $k+\pi$.
Consider a single band $|\psi(k)\rangle$ with periodicity $|\psi(k)\rangle=|\psi(k+2\pi)\rangle$. Then $\mathbf{P}T$ operator acts on the band as
\begin{equation}\label{psi}
  U_{PT}|\psi(k)\rangle^* = e^{i\phi(k)}|\psi(k+\pi)\rangle,
\end{equation}
where we expressed $\mathbf{P}T=U_{PT}K$ with $U_{PT}$ a unitary operator and $K$ the complex conjugation. From  Eq.~(\ref{PT2}), we have $U_{PT}U^*_{PT}=-1$, which leads to
\begin{equation}\label{phi}
  e^{i\phi(k+\pi)-i\phi(k)}=-1.
\end{equation}
Therefore, the PSA  establishes a connection between states at $k$ and $k+\pi$.

\begin{figure*}[t]
	\includegraphics[scale=1.0]{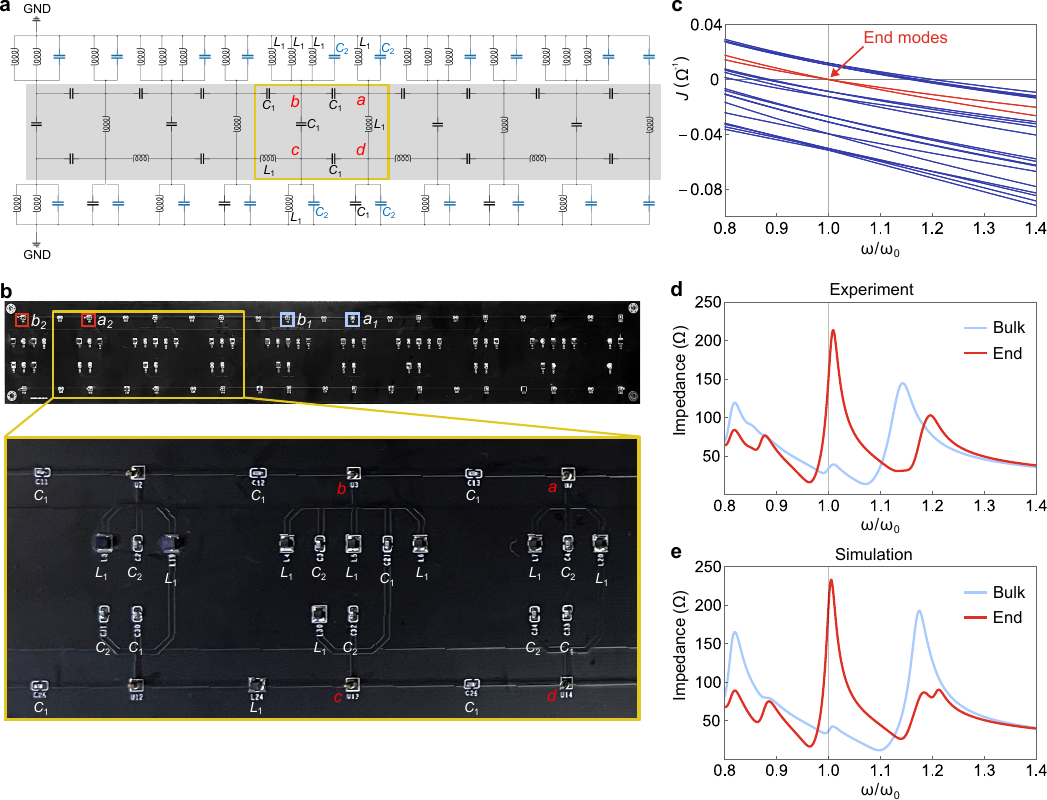}
	\caption{\textbf{Electric-circuit realization of the SSH model.} 
	\textbf{a},~Circuit diagram for the flux SSH model with five unit cells. The shadowed area faithfully simulates the flux SSH model. Each inductor (capacitor) corresponds to a negative (positive) hopping amplitude. A unit cell is indicated by the yellow rectangular.  The remaining part is designed to  tune the reference voltage of the electric circuit. 
	\textbf{b},~Photograph of our fabricated circuit board. 
	\textbf{c},~Theoretical spectrum of the circuit Laplacian as a function of the driving frequency. All frequency scales are normalized to the resonance frequency $\omega_0$. Two isolated modes crossing the gap, which correspond to  zero-energy eigenvalues of the circuit Laplacian at $ \omega=\omega_0 $, are marked in red. They correspond to the two topological end modes. 
	\textbf{d}, \textbf{e},~Experimental and simulated impedance responses versus normalized frequency. The end (bluk)  impedance is measured between $a_2$($a_1$) and $b_2$($a_2$) in \textbf{b}. The curve at the end (in the bulk) is marked in red (blue). The experimental curves well agree with the simulation curves, confirming the topological end states of the flux SSH model.
	 \label{fig:circuit}}
\end{figure*}	

Using Eqs.~(\ref{psi}) and (\ref{phi}), one can readily evaluate the Zak phase~\cite{Zak_phase}
\begin{equation}\label{zak}
	\gamma=\oint dk~\langle\psi(k) |i\partial_k|\psi(k)\rangle,
\end{equation}
by dividing the integration domain into two parts, $[-\pi,0]$ and $[0,\pi]$, and relating the two by $\mathbf{P}T$ (see Methods).  For a single band, one finds that the Zak phase is guaranteed to be $\gamma=\pi$. It must be pointed out that
the above analysis is completely general: We never used any details of the flux SSH model except the PSA in  (\ref{PT2}) and (\ref{PTL}) that the model satisfies. In other words, the result $\gamma=\pi$ is determined solely by PSA.

Now, given the system the flux dimerization, when will the Zak phase be zero? From the above discussion, this must occur in a configuration with a different PSA. For flux SSH model, this corresponds to the unit cell choice with inversion center at zero-flux plaquette [see Fig.~\ref{fig:tb}\textbf{b}]. One can easily check that in this case, although (\ref{PTL}) remains the same, PSA in (\ref{psi}) is changed to $(\mathbf{P}T)^2=+1$, which then dictates $\gamma=0$.

Thus , the two topological phases of flux SSH model directly corresponds to two distinct PSAs, with $(\mathbf{P}T)^2=\alpha\in \{\pm 1\}$, such that the Zak phase can be expressed as
\begin{equation}\label{eq:determined_gamma}
	\gamma=i\ln \alpha \mod 2\pi.
\end{equation}
This is in contrast to conventional cases, like the original SSH model, where symmetries [as in Eq.~(\ref{or})] cannot determine the topological phase.


Our claims above are confirmed by a direct calculation of the model (see Methods). Figure \ref{fig:tb}\textbf{d}-\textbf{f} shows the nontrivial case with
PSA \eqref{psi} and \eqref{phi}. The calculated band structure is plotted in Fig.~\ref{fig:tb}\textbf{d}, and we verifies that each band here has a $\pi$ Zak phase. This is another salient feature distinct from conventional systems where different bands are not guaranteed to have the same Zak phase.

A $\pi$ Zak phase requires the presence of 0D topological modes
at the end of the 1D chain. In our model, this occurs for  the first and the third bulk band gaps (the second gap is trivial since the Zak phases for the two bands below add up to zero). Such topological end modes are confirmed in Fig.~\ref{fig:tb}\textbf{e}, by our calculation of a chain with a finite length. The profiles of the two ingap states in the third gap is plotted in Fig.~\ref{fig:tb}\textbf{f}.

The proposed flux SSH model constitutes a minimal model for demonstrating the PSA determined topology. Below, we present experimental realization of this model in an electric-circuit array, which directly verifies our theory.

Electric circuits are governed by the Kirchhoff's law. In the frequency domain, it assumes the general form of
$I_i(\omega) = \sum_{j} J_{ij}(\omega) V_j(\omega)$, where $I_i(\omega)$ and $V_i(\omega)$ are the electric current and voltage at node $i$,
$J_{ij}(\omega)$ is the admittance between nodes $i$ and $j$, and the summation is over all adjacent nodes.
It follows that the behavior of a circuit is characterized by its $J(\omega)$ matrix, also known as the circuit Laplacian, and the task is to design a circuit whose $J(\omega)$ matrix can simulate the flux SSH model.

The design is actually very straightforward, thanks to the characters of capacitors and inductors which naturally exhibit a phase difference in their responses~\cite{Topo_circuit_1,Topo_circuit_2,Ronny_NP_2018}. As shown in Fig.~\ref{fig:circuit}\textbf{a}, one just needs to use capacitor ($C_1$) for the blue bond and inductor ($L_1$) for the red bond. Then a unit cell contains four capacitors and two inductors (and four nodes). The additional inductors and capacitors (with values $L_1$ and $C_2$) at top and bottom in Fig.~\ref{fig:circuit}\textbf{a} are used to facilitate measurement. They are not essential to PSA and topology.

For this simple circuit array, one readily derives that
\begin{equation}
  J(\omega, k)=i\omega C_1 H(\omega, k),
\end{equation}
and at driving frequency $\omega=\omega_0\equiv 1/\sqrt{L_1C_1}$,
\begin{equation}\label{HH}
  H(\omega_0,k)=\mathcal{H}_\text{fSSH}(k)-\lambda \mathds{1},
\end{equation}
where $\mathcal{H}_\text{fSSH}$ is just the flux SSH model put in dimensionless form (see Methods), and $\lambda=C_2/C_1$
is shift that can be utilized to probe the topological end mode.

The measurement is on the impedance response $Z_{ab}(\omega)$ between two nodes $a$ and $b$, which can be expressed as
\begin{equation}
	Z_{ab}(\omega)=\frac{V_{a}-V_{b}}{I_{ab}}=\sum_{n}\frac{ \vert \psi _{n,a}-\psi_{n,b} \vert ^{2}}{j_{n}(\omega)},
\end{equation}
where $j_{n}$ and $\{\psi_{n,i}\}$ are the $n$-th eigenvalue and eigenmode of $J(\omega)$.
In $Z_{ab}(\omega)$, the mode with $j_{n}$ close to zero (called the zero-admittance mode) will dominate the response.
For any target mode, we can utilize the $\lambda$ term in (\ref{HH}) (by tuning $C_2$) to shift its eigenvalue to zero. Here, we focus on the topological end mode corresponding to the one in the third gap of $\mathcal{H}_\text{fSSH}$ in Fig.~\ref{fig:ssh}\textbf{e}, which has a value of $\sim 1.48$, so we choose $\lambda=1.48$ in our design. This makes the topological end mode the zero-admittance mode at frequency $\omega=\omega_0$. In Fig.~\ref{fig:circuit}\textbf{c}, we show the simulated $J(\omega)$ spectrum of our designed circuit, which confirms this point.


Experimentally, we fabricate the designed circuit on a printed circuit board, as shown in Fig.~\ref{fig:circuit}\textbf{b}. It has a length of five unit cells, not long but sufficient to discern the topological end modes. We perform impedance measurement on two pairs of nodes. The first pair $a_1$ and $b_1$ are in the bulk (the middle cell), and the second pair $a_2$ and $b_2$ are at the end (the first cell), as indicated in Fig.~\ref{fig:circuit}\textbf{b}. From the above analysis, one expects that the topological end mode of flux SSH model should manifest as a peak at
$\omega=\omega_0$ for measurement at $a_2$ and $b_2$, and this peak will disappear for measurement at $a_1$ and $b_1$. 
This is confirmed by the measured results in Fig.~\ref{fig:circuit}\textbf{d}. The experimental curves also agree very well with results from numerical simulations of the circuit (see Fig.~\ref{fig:circuit}\textbf{e}).

In summary, we have discovered an extraordinary phenomenon beyond the common wisdom regarding topological phases, i.e., the projective algebraic structure of symmetries can completely determine the topological phase. We propose a simple model, the flux SSH model, which demonstrates the phenomenon. In the present case, every band is enforced by PSA to have nontrivial Zak phase. We also provide the first experimental proof of this remarkable phenomenon using a designed electric circuit array. 
Considering the rich crystal space group symmetries, we expect that there will be an abundance of such intriguing effect to be discovered for PSAs. Our proposed flux dimerization may serve as an effective design approach to realize novel PSA determined topological phases, applying to a wide range of physical systems besides electric circuits, such as cold atoms~\cite{Topo_cold_atoms,Atificial_Gauge_RMP}, phononic/photonic crystals~\cite{Haldane_Photonics,Topological_photonics,Baile_Topo_Acoustics,Haoran_NRM_2023}, mechanical networks~\cite{kane2014topological,Ma_topological_2019}, and etc.

%

\bigskip
\def\bibsection{\ } 
\noindent \textbf{REFERENCES}
\bibliographystyle{naturemag}
\bibliography{reference}

\begin{thebibliography}{10}
\expandafter\ifx\csname url\endcsname\relax
  \def\url#1{\texttt{#1}}\fi
\expandafter\ifx\csname urlprefix\endcsname\relax\def\urlprefix{URL }\fi
\providecommand{\bibinfo}[2]{#2}
\providecommand{\eprint}[2][]{\url{#2}}

\bibitem{Kane_RMP}
\bibinfo{author}{Hasan, M.~Z.} \& \bibinfo{author}{Kane, C.~L.}
\newblock \bibinfo{title}{Colloquium: Topological insulators}.
\newblock \emph{\bibinfo{journal}{Rev. Mod. Phys.}} \textbf{\bibinfo{volume}{82}}, \bibinfo{pages}{3045--3067} (\bibinfo{year}{2010}).

\bibitem{XLQi_RMP}
\bibinfo{author}{Qi, X.-L.} \& \bibinfo{author}{Zhang, S.-C.}
\newblock \bibinfo{title}{Topological insulators and superconductors}.
\newblock \emph{\bibinfo{journal}{Rev. Mod. Phys.}} \textbf{\bibinfo{volume}{83}}, \bibinfo{pages}{1057--1110} (\bibinfo{year}{2011}).

\bibitem{Schnyder_RMP}
\bibinfo{author}{Chiu, C.-K.}, \bibinfo{author}{Teo, J. C.~Y.}, \bibinfo{author}{Schnyder, A.~P.} \& \bibinfo{author}{Ryu, S.}
\newblock \bibinfo{title}{Classification of topological quantum matter with symmetries}.
\newblock \emph{\bibinfo{journal}{Rev. Mod. Phys.}} \textbf{\bibinfo{volume}{88}}, \bibinfo{pages}{035005} (\bibinfo{year}{2016}).

\bibitem{SSH}
\bibinfo{author}{Su, W.~P.}, \bibinfo{author}{Schrieffer, J.~R.} \& \bibinfo{author}{Heeger, A.~J.}
\newblock \bibinfo{title}{Solitons in polyacetylene}.
\newblock \emph{\bibinfo{journal}{Phys. Rev. Lett.}} \textbf{\bibinfo{volume}{42}}, \bibinfo{pages}{1698--1701} (\bibinfo{year}{1979}).

\bibitem{Zak_phase}
\bibinfo{author}{Zak, J.}
\newblock \bibinfo{title}{Berry's phase for energy bands in solids}.
\newblock \emph{\bibinfo{journal}{Phys. Rev. Lett.}} \textbf{\bibinfo{volume}{62}}, \bibinfo{pages}{2747--2750} (\bibinfo{year}{1989}).

\bibitem{Zhao_PT_cls}
\bibinfo{author}{Zhao, Y.~X.}, \bibinfo{author}{Schnyder, A.~P.} \& \bibinfo{author}{Wang, Z.~D.}
\newblock \bibinfo{title}{Unified theory of $pt$ and $cp$ invariant topological metals and nodal superconductors}.
\newblock \emph{\bibinfo{journal}{Phys. Rev. Lett.}} \textbf{\bibinfo{volume}{116}}, \bibinfo{pages}{156402} (\bibinfo{year}{2016}).

\bibitem{PhysRevB.102.161117}
\bibinfo{author}{Zhao, Y.~X.}, \bibinfo{author}{Huang, Y.-X.} \& \bibinfo{author}{Yang, S.~A.}
\newblock \bibinfo{title}{{${\mathbb{Z}}_{2}$}-projective translational symmetry protected topological phases}.
\newblock \emph{\bibinfo{journal}{Phys. Rev. B}} \textbf{\bibinfo{volume}{102}}, \bibinfo{pages}{161117} (\bibinfo{year}{2020}).

\bibitem{PT_Switch_PRL}
\bibinfo{author}{Zhao, Y.~X.}, \bibinfo{author}{Chen, C.}, \bibinfo{author}{Sheng, X.-L.} \& \bibinfo{author}{Yang, S.~A.}
\newblock \bibinfo{title}{Switching spinless and spinful topological phases with projective $pt$ symmetry}.
\newblock \emph{\bibinfo{journal}{Phys. Rev. Lett.}} \textbf{\bibinfo{volume}{126}}, \bibinfo{pages}{196402} (\bibinfo{year}{2021}).

\bibitem{Xue_prl_2022}
\bibinfo{author}{Xue, H.} \emph{et~al.}
\newblock \bibinfo{title}{Projectively enriched symmetry and topology in acoustic crystals}.
\newblock \emph{\bibinfo{journal}{Phys. Rev. Lett.}} \textbf{\bibinfo{volume}{128}}, \bibinfo{pages}{116802} (\bibinfo{year}{2022}).

\bibitem{Qiu_PRL_2022}
\bibinfo{author}{Li, T.} \emph{et~al.}
\newblock \bibinfo{title}{Acoustic m\"obius insulators from projective symmetry}.
\newblock \emph{\bibinfo{journal}{Phys. Rev. Lett.}} \textbf{\bibinfo{volume}{128}}, \bibinfo{pages}{116803} (\bibinfo{year}{2022}).

\bibitem{PT_Expert_PRL}
\bibinfo{author}{Meng, Y.} \emph{et~al.}
\newblock \bibinfo{title}{Spinful topological phases in acoustic crystals with projective $pt$ symmetry}.
\newblock \emph{\bibinfo{journal}{Phys. Rev. Lett.}} \textbf{\bibinfo{volume}{130}}, \bibinfo{pages}{026101} (\bibinfo{year}{2023}).

\bibitem{chen2023classification}
\bibinfo{author}{Chen, Z.~Y.}, \bibinfo{author}{Zhang, Z.}, \bibinfo{author}{Yang, S.~A.} \& \bibinfo{author}{Zhao, Y.~X.}
\newblock \bibinfo{title}{Classification of time-reversal-invariant crystals with gauge structures}.
\newblock \emph{\bibinfo{journal}{Nature Communications}} \textbf{\bibinfo{volume}{14}}, \bibinfo{pages}{743} (\bibinfo{year}{2023}).

\bibitem{Jonah_PRL_2023}
\bibinfo{author}{Herzog-Arbeitman, J.}, \bibinfo{author}{Song, Z.-D.}, \bibinfo{author}{Elcoro, L.} \& \bibinfo{author}{Bernevig, B.~A.}
\newblock \bibinfo{title}{Hofstadter topology with real space invariants and reentrant projective symmetries}.
\newblock \emph{\bibinfo{journal}{Phys. Rev. Lett.}} \textbf{\bibinfo{volume}{130}}, \bibinfo{pages}{236601} (\bibinfo{year}{2023}).

\bibitem{ZhangChen_prl_2023}
\bibinfo{author}{Zhang, C.}, \bibinfo{author}{Chen, Z.~Y.}, \bibinfo{author}{Zhang, Z.} \& \bibinfo{author}{Zhao, Y.~X.}
\newblock \bibinfo{title}{General theory of momentum-space nonsymmorphic symmetry}.
\newblock \emph{\bibinfo{journal}{Phys. Rev. Lett.}} \textbf{\bibinfo{volume}{130}}, \bibinfo{pages}{256601} (\bibinfo{year}{2023}).

\bibitem{Wigner_Projective}
\bibinfo{author}{Wigner, E.}
\newblock \bibinfo{title}{On unitary representations of the inhomogeneous lorentz group}.
\newblock \emph{\bibinfo{journal}{Annals of Mathematics}} \textbf{\bibinfo{volume}{40}}, \bibinfo{pages}{149--204} (\bibinfo{year}{1939}).

\bibitem{Topo_circuit_1}
\bibinfo{author}{Ningyuan, J.}, \bibinfo{author}{Owens, C.}, \bibinfo{author}{Sommer, A.}, \bibinfo{author}{Schuster, D.} \& \bibinfo{author}{Simon, J.}
\newblock \bibinfo{title}{Time- and site-resolved dynamics in a topological circuit}.
\newblock \emph{\bibinfo{journal}{Phys. Rev. X}} \textbf{\bibinfo{volume}{5}}, \bibinfo{pages}{021031} (\bibinfo{year}{2015}).

\bibitem{Topo_circuit_2}
\bibinfo{author}{Albert, V.~V.}, \bibinfo{author}{Glazman, L.~I.} \& \bibinfo{author}{Jiang, L.}
\newblock \bibinfo{title}{Topological properties of linear circuit lattices}.
\newblock \emph{\bibinfo{journal}{Phys. Rev. Lett.}} \textbf{\bibinfo{volume}{114}}, \bibinfo{pages}{173902} (\bibinfo{year}{2015}).

\bibitem{Ronny_NP_2018}
\bibinfo{author}{Imhof, S.} \emph{et~al.}
\newblock \bibinfo{title}{Topolectrical-circuit realization of topological corner modes}.
\newblock \emph{\bibinfo{journal}{Nature Physics}} \textbf{\bibinfo{volume}{14}}, \bibinfo{pages}{925--929} (\bibinfo{year}{2018}).

\bibitem{Topo_cold_atoms}
\bibinfo{author}{Cooper, N.~R.}, \bibinfo{author}{Dalibard, J.} \& \bibinfo{author}{Spielman, I.~B.}
\newblock \bibinfo{title}{Topological bands for ultracold atoms}.
\newblock \emph{\bibinfo{journal}{Rev. Mod. Phys.}} \textbf{\bibinfo{volume}{91}}, \bibinfo{pages}{015005} (\bibinfo{year}{2019}).

\bibitem{Atificial_Gauge_RMP}
\bibinfo{author}{Dalibard, J.}, \bibinfo{author}{Gerbier, F.}, \bibinfo{author}{Juzeli\ifmmode~\bar{u}\else \={u}\fi{}nas, G.} \& \bibinfo{author}{\"Ohberg, P.}
\newblock \bibinfo{title}{Colloquium: Artificial gauge potentials for neutral atoms}.
\newblock \emph{\bibinfo{journal}{Rev. Mod. Phys.}} \textbf{\bibinfo{volume}{83}}, \bibinfo{pages}{1523--1543} (\bibinfo{year}{2011}).

\bibitem{Haldane_Photonics}
\bibinfo{author}{Raghu, S.} \& \bibinfo{author}{Haldane, F. D.~M.}
\newblock \bibinfo{title}{Analogs of quantum-hall-effect edge states in photonic crystals}.
\newblock \emph{\bibinfo{journal}{Phys. Rev. A}} \textbf{\bibinfo{volume}{78}}, \bibinfo{pages}{033834} (\bibinfo{year}{2008}).

\bibitem{Topological_photonics}
\bibinfo{author}{Ozawa, T.} \emph{et~al.}
\newblock \bibinfo{title}{Topological photonics}.
\newblock \emph{\bibinfo{journal}{Rev. Mod. Phys.}} \textbf{\bibinfo{volume}{91}}, \bibinfo{pages}{015006} (\bibinfo{year}{2019}).

\bibitem{Baile_Topo_Acoustics}
\bibinfo{author}{Yang, Z.} \emph{et~al.}
\newblock \bibinfo{title}{Topological acoustics}.
\newblock \emph{\bibinfo{journal}{Phys. Rev. Lett.}} \textbf{\bibinfo{volume}{114}}, \bibinfo{pages}{114301} (\bibinfo{year}{2015}).

\bibitem{Haoran_NRM_2023}
\bibinfo{author}{Xue, H.}, \bibinfo{author}{Yang, Y.} \& \bibinfo{author}{Zhang, B.}
\newblock \bibinfo{title}{Topological acoustics}.
\newblock \emph{\bibinfo{journal}{Nature Reviews Materials}} \textbf{\bibinfo{volume}{7}}, \bibinfo{pages}{974--990} (\bibinfo{year}{2022}).

\bibitem{kane2014topological}
\bibinfo{author}{Kane, C.~L.} \& \bibinfo{author}{Lubensky, T.~C.}
\newblock \bibinfo{title}{Topological boundary modes in isostatic lattices}.
\newblock \emph{\bibinfo{journal}{Nature Physics}} \textbf{\bibinfo{volume}{10}}, \bibinfo{pages}{39--45} (\bibinfo{year}{2014}).

\bibitem{Ma_topological_2019}
\bibinfo{author}{Ma, G.}, \bibinfo{author}{Xiao, M.} \& \bibinfo{author}{Chan, C.~T.}
\newblock \bibinfo{title}{Topological phases in acoustic and mechanical systems}.
\newblock \emph{\bibinfo{journal}{Nature Reviews Physics}} \textbf{\bibinfo{volume}{1}}, \bibinfo{pages}{281--294} (\bibinfo{year}{2019}).

\end{thebibliography}

\bigskip
\noindent \textbf{\large Methods}\\
\noindent \textbf{Gauge invariance of PSA} \\
One may modify $\mathbf{P}T$ by an arbitrary phase $e^{i\theta}$, as $\mathbf{P}T\mapsto e^{i\theta}\mathbf{P}T$. One finds that
\begin{equation}
	\begin{split}
	(e^{i\theta}\mathbf{P}T)^2&=	e^{i\theta}\mathbf{P}Te^{i\theta}\mathbf{P}T\\ &=e^{i\theta}e^{-i\theta}(\mathbf{P}T)^2=(\mathbf{P}T)^2.
	\end{split}
\end{equation}
In the second equality, we have used the fact that $\mathbf{P}T$ is antiunitary, and therefore a complex number is conjugated after commuting with $\mathbf{P}T$. Hence, we see $(\mathbf{P}T)^2=\alpha$ is gauge invariant. It is known as a cohomology invariant for the PSA. Similarly, one can verify the invariance of (\ref{PTL}) under $\mathbf{P}T\mapsto e^{i\theta}\mathbf{P}T$ and $L \mapsto e^{i\theta'}L$. 
%
\\
\\
\noindent \textbf{Derivation of the enforced Zak phase} \\
The integrand in Eq.~(\ref{zak}) is known as the Berry connection $\mathcal{A}(k)$.  From Eq.~(\ref{psi}), we find that
\begin{equation}
  |\psi(k+\pi)\rangle = e^{-i\phi(k)} U_{PT}|\psi(k)\rangle^*.
\end{equation}
Substituting this into the Berry connection expression at $k+\pi$, we obtain
\begin{equation}
	\mathcal{A}(k +  \pi) = -\mathcal{A}(k) + \partial_{k}\phi(k).
\end{equation}
Hence, the Zak phase can be expressed as
\begin{equation}
	\gamma = \int_{0}^\pi dk~(\mathcal{A}(k) + \mathcal{A}(k+ \pi) ) = \phi(\pi)-\phi(0).
\end{equation}
Then, using Eq.~(\ref{phi}), we find that $\gamma=\pi$. 

In addition, following similar derivation, one finds that for PSA with $(\mathbf{P}T)^2=\alpha\in \{\pm 1\}$ and $(\mathbf{P}T)L(\mathbf{\mathbf{P}}T)^{-1}=-L^{-1}$, we must have
\begin{equation}
  \gamma =i\ln \alpha \mod 2\pi,
\end{equation}
which is Eq.~\eqref{eq:determined_gamma} in the main text. 
\\
\\
\textbf{Flux SSH model}\\
Our proposed flux SSH model is a minimal model that realizes the PSA in (\ref{PT2}) and (\ref{PTL}). 
According to Fig.~\ref{fig:tb}\textbf{a}, the explicit form of the model in real space can be written as 
\begin{equation}\label{eq:hami1}
	\begin{aligned}
	{H} = t\sum_{i} ( c^\dagger_{i,a}  c_{i,b} +  c^\dagger_{i,b}  c_{i,c} +  c^\dagger_{i,c}  c_{i,d}  -  c^\dagger_{i,d}  c_{i,a}) \\
	+ t\sum_{i} ( c^\dagger_{i+1,b}  c_{i,a} -  c^\dagger_{i+1,c}  c_{i,d}) + \rm{h.c.},
\end{aligned}
\end{equation}
where the hopping parameter $t$ is taken to be real positive, $i$ labels the unit cell, the first term is intracell hopping, and the second term is intercell hopping. For the minimal model, we include the nearest neighbor hopping. 
One can certainly add more complicated terms, such as far-neighboring hoppings, but as long as they respect the PSA, the topological character of the system must remain unchanged. 

Transforming to $k$ space, the flux SSH model takes the form of
\begin{eqnarray}\label{eq:hami2}
	\mathcal{H}_\text{fSSH} = \begin{bmatrix}
		0 & 1+e^{ik} & 0 & - 1 \\
		1+e^{-ik} & 0 &  1 & 0 \\
		0 &  1 & 0 & 1 - e^{-ik} \\
		- 1 & 0 & 1 - e^{ik} & 0
	\end{bmatrix},
\end{eqnarray}
where we put it in dimensionless form (in unit of $t$). Its band structure consists of four bands with 
\begin{equation}
  E(k)= \pm \sqrt{3\pm2 \sqrt{1+ \cos^2k}},
\end{equation}
which has been plotted in Fig.~\ref{fig:tb}\textbf{d}. Direct calculation shows that every band here carries a $\pi$ Zak phase, as we predicted based on PSA. The topological end modes for a flux SSH chain with 25 unit cells have been confirmed by the results in Fig.~\ref{fig:tb}\textbf{e} and \textbf{f}.

In connection with the original SSH model, we also provide the following intuitive picture. 
In the flux SSH model, consider the coupling between two neighboring sites along the chain, e.g., between sites $a$ and $b$. 
One can readily identify two leading interaction paths, one is direct hopping between the two, the other is to go around three edges of a plaquette. Then, the effective hopping amplitude is the superposition of two paths. Clearly, the coupling is enhanced for $0$-flux plaquettes and reduced for $\pi$-flux plaquettes. As a result, one can imagine that the effective hopping amplitudes form a dimerization pattern similar to the original SSH model. This offers an intuitive understanding of the result. \\

\noindent \textbf{The spacetime inversion operator} \\
For the flux SSH model with $\mathcal{H}_{\text{fSSH}}(k)$ in Eq.~\eqref{eq:hami2}, the unitary matrix associated to $\mathbf{P}T$ is given by
\begin{eqnarray}
	U_{PT} & = 
	\begin{bmatrix} 
		0 & 0 & -1 & 0 \\
		0 & 0 & 0 & -1 \\
		1 & 0 & 0 & 0 \\
		0 & 1 & 0 & 0
	\end{bmatrix} , 
\end{eqnarray}
which satisfies
\begin{eqnarray}
	U_{PT} U_{PT}^{*}= -1 .
\end{eqnarray}
It is straightforward to verify that
\begin{eqnarray}
	U_{PT} \mathcal{H}_{\text{fSSH}}^*(k) U_{PT}^\dagger=\mathcal{H}_{\text{fSSH}}(k+\pi) ,
\end{eqnarray}
i.e., $\bm{\mathrm{P}}T$ translates $k$ by $\pi$.\\

\noindent \textbf{ $J$ matrix for the flux SSH circuit}\\
By Kirchhoff's law, our designed circuit has its circuit Laplacian
$J(\omega,k)=i\omega C_1 H(\omega,k)$, where
\begin{equation*}
	H(\omega,k)=\begin{bmatrix}
			2\eta-\lambda & 1+e^{ik} & 0 & -\zeta \\
			1+ e^{-ik} & 3\eta-\lambda & 1 & 0 \\
			0 & 1 & 2\eta-\lambda & 1 - \zeta e^{-ik} \\
			-\zeta & 0 & 1 - \zeta e^{ik} & 2\eta-\lambda
		\end{bmatrix} ,
\end{equation*}
where $\eta=(L_1C_1)^{-1}/\omega^{2}-1$, $\zeta=(L_1C_1)^{-1}/\omega^{2}$, and $\lambda=C_2/C_1$.  
When we tune the driving frequency $\omega$ to the LC resonance frequency $\omega_0=1/\sqrt{ L_1 C_1}$, we have $\eta=0$ and $\zeta=1$. Compared with the flux SSH model in Eq.~(\ref{eq:hami2}), we immediately notice that 
\begin{equation}\label{eq: Hc_hal}
	H(\omega_0,k)=\mathcal{H}_\text{fSSH}(k)-\lambda\mathds{1}.
\end{equation}
Therefore, the designed circuit constitutes a realization of our proposed flux SSH model. 
\\
\\
\noindent \textbf{Experimental details} \\
In the designed circuit, we choose $C_1 = 1$ nF, $ L_1 = 5.6$ $\mathrm{\mu H}$, so the resonant frequency  is $2.1268$ $\mathrm{MHz}$. Ideally, a single capacitor with capacitance $1.48$ nF can be chosen to realize the desired $C_2$, but in practice, it is difficult to find proper capacitors with this exact value. Therefore, we use two capacitors whose capacitances are $ C_{2a} = 1$ $\mathrm{nF}$ and $ C_{2b} = 0.47$ $\mathrm{nF}$ in parallel to realize $C_2$. In this way, the capacitance $C_2$ achieved is $ 1.47$ $\mathrm{nF}$, slightly lower than the ideal value. 

In the electric circuit that we fabricated, the part number of $ C_1 $ and $C_{2a}$ is GRM1885C1H102FA01D, the part number of $ L_1 $ is SLF10145T-5R6M3R2-PF, and the part number of $ C_{2b} $ is GRM1885C1H471FA01D. All impedance measurements were performed with a HP 4194A Impedance/Gain-Phase Analyzer.
\\
\\

\noindent \textbf{\large Data availability}\\
The data that support the plots within this paper and other findings of this study are available from the corresponding author upon reasonable request.\\

\noindent \textbf{\large Acknowledgements}\\
This work is supported by National Natural Science Foundation of China (Grants No.~12161160315 and No.~12174181), Basic Research Program of Jiangsu Province (Grant No.~BK20211506). \\

\noindent \textbf{\large Author contributions}
Y.X.Z. conceived the idea. Z.Y.C. and Y.X.Z. developed the theory. G.J. and S.J.Y. constructed the electric-circuit model and performed numerical calculations. G.J. and X.M.Z. conducted the experiment. G.J., Z.Y.C., W.B.R., S.A.Y. and Y.X.Z. wrote the manuscript with input from all authors. S.A.Y and Y.X.Z. supervised the project.\\

\noindent \textbf{\large Competing interests}
The authors declare no competing interests.

\end{document}